\documentclass[twocolumn,showpacs,preprintnumbers,superscriptaddress,amsmath,floatfix,amssymb,prd]{revtex4}
\usepackage{graphicx}
\usepackage{amsfonts,amsmath,amsxtra}

\newcommand{\beq}[1]{\begin{eqnarray}\label{#1}}
\newcommand\eeq {\end{eqnarray}}
\newcommand\bqa {\begin{eqnarray}}
\newcommand\eqa {\end{eqnarray}}

\newcommand{\bear}{\begin{array}}
\newcommand{\enar}{\end{array}}

\newcommand{\R}{\mathbb{R}}
\newcommand{\C}{\mathbb{C}}
\newcommand{\K}{\mathcal{K}}

\newcommand{\Z}{\mathbb{Z}}

\newcommand{\N}{\mathbb{N}}

\begin{document}
\def\le{\langle}
\def\re{\rangle}
\def\dg{^{\dag}}
\def\K{{\cal K}}
\def\n{{\cal N}}
\font\maj=cmcsc10 \font\itdix=cmti10
\def\N{\mbox{I\hspace{-.15em}N}}
\def\H{\mbox{I\hspace{-.15em}H\hspace{-.15em}I}}
\def\1{\mbox{I\hspace{-.15em}1}}
\def\pN{\rm{I\hspace{-.1em}N}}
\def\Z{\mbox{Z\hspace{-.3em}Z}}
\def\pZ{{\rm Z\hspace{-.2em}Z}}
\def\R{{\rm I\hspace{-.15em}R}}
\def\pR{{\rm I\hspace{-.10em}R}}
\def\C{\hspace{3pt}{\rm l\hspace{-.47em}C}}
\def\Q{\mbox{l\hspace{-.47em}Q}}
\def\b{\begin{equation}}
\def\e{\end{equation}}
\def\bee{\begin{enumerate}}
\def\wt{\widetilde}

\title{Large Angular Scale CMB Anisotropy from an Excited Initial Mode}
\author{A. Sojasi}
\email{sojasi@iaurasaht.ac.ir} \affiliation{Department of Physics,
Rasht Branch, Islamic Azad University, Rasht, Guilan, Iran}
\author{M. Mohsenzadeh}
\email{mohsenzadeh@qom-iau.ac.ir} \affiliation{Department of
Physics, Qom Branch, Islamic Azad University, Qom, Iran}
\author{E. Yusofi}
\email{e.yusofi@iauamol.ac.ir} \affiliation{Department of Physics,
Ayatollah Amoli Branch, Islamic Azad University, Amol, Mazandaran,
Iran}
\date{\today}

\begin{abstract}
According to inflationary cosmology, the CMB anisotropy gives
an opportunity to test predictions of new physics hypotheses.
The initial state of quantum fluctuations is one of the important
options at high energy scale, as it can affect observables such
as the CMB power spectrum. In this study a quasi-de Sitter
inflationary background with approximate de Sitter mode function
built over the Bunch-Davies mode is applied to investigate the
scale-dependency of the CMB anisotropy. The
recent Planck constraint on spectral index motivated us to
examine the effect of a new excited mode function (instead of pure
de Sitter mode) on the CMB anisotropy at large angular scales. In
so doing, it is found that the angular scale-invariance in the
CMB temperature fluctuations is broken and in the limit $ \ell<200
$ a tiny deviation appears. Also, it is shown that the
power spectrum of CMB anisotropy is dependent on a free
parameter with mass dimension $H<<M_{*}<M_{p}$ and on the slow-roll
parameter $\epsilon $.

\end{abstract}
Submitted to Chinese Physics C


\maketitle
\section{Introduction}

Today's cosmological observations give us a useful pattern of
the Cosmic Microwave Background Anisotropies (CMBA). This pattern
gives us a good snapshot of the temperature fluctuations of
photons on the last scattering surface (LSS) \cite{c1, c2}.
Temperature anisotropies in the photons arise due to several
physical effects. One of the interesting effects in which the CMB
photons are gravitationally red-shifted on the LSS, as they
decoupled from matter, is known as the Sachs-Wolfe effect. Moreover,
it is the dominant effect in the CMB anisotropy at large angular
scales \cite{c1, c2, c3}. Indeed, this anisotropy is seeded by
the primordial perturbation in the early universe which manifests
itself in the anisotropies of the CMB photons as well as the
matter density perturbation today \cite{c2, c3, c4, c5, c6}. So,
the large angular
scale anisotropies of these photons are actually encoded by primordial perturbations.\\
It is well-known that a particular pattern of thermal anisotropies depends not only on the particular model of inflation, but also on the initial vacuum state of the quantum fluctuations [7-16]. It is usually considered that when these fluctuations are generated, they are initially in the minimum energy state so-called Bunch-Davies or de Sitter vacuum \cite{c13,c14,c15,c151,c16}. But according to new observational data about the scalar spectral index, it has been shown that the background geometry of inflation is not pure de Sitter \cite{c7,c8}. So, one can consider a non-Bunch-Davies state as an initial state of scalar field fluctuations of which the physical origin is unknown. According to some studies, it may be due to some initial condition arising from pre-inflationary evolution as calculated in \cite{c17}; or from a nonsingular bounce as studied in \cite{c18,c19,c20,c21,c211,c212,c213,c214}; or from trans-Planckian physics in \cite{c22,c23}; or from the string theory effects in \cite{c24,c25,c26}; and so on. These motivate us to choose, from among the various possibilities for initial vacuum states of quantum fluctuations, an excited state which is constructed based on an excited de Sitter mode and which leads to the higher order trans-Planckian corrections \cite{c27}. Actually, these higher order corrections are produced because the applied non-trivial mode are non-linear (up to second order) with respect to $ \frac{1}{k\eta} $ \cite{c27,c28,c29}. On the other hand the free parameter with mass dimension as a cutoff scale in the corrections emphasises that the effect of these excited modes is bounded. In addition, as a result of using this excited initial state, it has been shown that we can have particle creation~\cite{c30}. They have also been used to calculate the scale-dependency of the primordial power spectrum \cite{c31}. In this work, it is expected that employing an initial excited de Sitter mode for calculating the primordial perturbed gravitational potential leads to the breaking of angular-scale symmetry in the CMBA power spectrum.\\
The layout of the paper is as follows. In Section 2, the excited de
Sitter mode for nearly de Sitter inflationary background is
briefly introduced and the power spectrum with this mode is
calculated. In Section 3, the definition of large-scale
inhomogeneity and anisotropy is reviewed and the scale-dependency
of large-scale matter density perturbation and Large-Angular Scale
CMB Anisotropy resulting from the excited de Sitter mode are
calculated. Conclusions are given in the final section.

\section{Non-trivial Initial Vacuum for Inflationary Background}
In curved space-time, the quantization of a scalar field is
similar to the quantization in flat space-time (i.e. Minkowskian),
but the gravitational interaction due to the curvature of
space-time can act as an external classical field on flat
space-time, which could be generally non-homogeneous and
non-stationary \cite{c32}. In general, due to the absence of
Killing vector in a curved space-time, the notion of vacuum is
ambiguous \cite{c33}. However, in de Sitter space with
maximal symmetry, it is possible to define the vacuum state
under the de Sitter symmetry group~\cite{c16}. So, it is a logical
choice that we consider de Sitter space as a background for our
theory and use the following metric to describe the expanding
inflationary universe with curved space-time \b
\label{equ1}ds^{2}=dt^{2}-a^{2}(t){d\textbf{x}}^{2}=a^{2}(\eta)({d\eta}^2-{d\textbf{x}}^2).
\e For de Sitter space-time, the scale factor is given by
$a(t)=\exp(Ht)$, or equivalently in a conformal formalism
$a(\eta)=-\frac{1}{H\eta}$. There are several models for the
inflationary universe, but the simple and most popular one is a
minimally coupled scalar field (inflaton) in an inflating
background \b \label{equ2}S=\frac{1}{2}\int
d^4x\sqrt{-g}\Big(R-(\nabla \varphi)^2-m^2\varphi^2\Big). \e The
Fourier components corresponding to the inflaton field satisfy the
equation of motion \cite{c5}\b
\label{equ3}{\varphi''}_{k}-\frac{2}{\eta}{\varphi'}_{k}+(k^{2}+a^2m^2)\varphi_{k}=0,
 \e
where the prime denotes the derivative with respect to conformal
time $\eta$. By considering the massless case and re-scaling of
$\varphi_{k}$ as $u_{k}=a\varphi_{k}$, Equation (\ref{equ3})
becomes \b
 \label{equ4} {u''}_{k}+\omega_{k}^{2}(\eta)u_{k}=0,
  \e
where $ \omega_{k}^{2}(\eta)=k^{2}-\frac{a''}{a} $. The solutions of
this equation give the positive and negative frequency modes. So
the general solutions of the equation of motion are given by \b
\label{equ5}u_{k}=A_{k}H_{\mu}^{(1)}(|k\eta|)+B_{k}H_{\mu}^{(2)}(|k\eta|),
 \e
 where $ H_{\mu}^{(1, 2)} $ are the Hankel functions of the first and second kind. By imposing $ [u(x,\eta),\pi(y,\eta)]=i\delta(x-y) $, namely the equal-time commutation relations, and also implementing secondary quantization in Fock representation, the canonical quantization can be performed for the scalar field $ u $ and its canonically conjugate momentum $ \pi\equiv u' $, \cite{c33}.

\subsection{Excited de Sitter Mode for Quasi-de Sitter Space-time}
  In the pure de Sitter background, with $
\frac{a^{''}}{a}=\frac{2}{\eta^{2}}$, the general solution of
(\ref{equ4}) becomes [5]\b \label{equ6}
u_{k}^{dS}=\frac{A_{k}}{\sqrt{2k}}\Big(1-\frac{i}{k\eta}\Big)e^{-ik\eta}+\frac{B_{k}}{\sqrt{2k}}\Big(1+\frac{i}{k\eta}\Big)e^{+ik\eta},
\e where $B_{k}$ and $A_{k}$ are Bogoliubov coefficients. By
setting $A_{k}=1$  and $B_{k}=0$, this solution leads to the
Bunch-Davies mode \cite{c16} \b \label{equ7}
u_{k}^{BD}=\frac{1}{\sqrt{2k}}\Big(1-\frac{i}{k\eta}\Big)e^{-ik\eta}.\e
As previously mentioned, the observational data released by
WMAP and Planck reveal that the inflationary universe can be
described by \emph{nearly de Sitter space-time}, where for the
initial state any \emph{approximate de Sitter mode} can be
selected as an acceptable mode. Motivated by this fact, in
Ref. [33] we suggested the following excited mode function as the
fundamental mode during the inflationary era:
  \b \label{equ8}
u_{k}^{exc}\simeq\frac{1}{\sqrt{2k}}\Big(1-\frac{i}{k\eta}-\frac{1}{2}(\frac{i}{k\eta})^{2}\Big)e^{-ik\eta}.\e
In the far past time limit, in a straightforward manner, if
$\frac{1}{k\eta}\neq0$ and
$\frac{1}{k^{2}\eta^{2}}\rightarrow{0}$, this mode function
asymptotically approaches to the de Sitter mode function.\\
Our reason to introduce this excited-de Sitter solution are
physical considerations, as noted in Section 3 of Ref. [32]. In
\cite{c28}, for the first time, we used this excited solution with
the auxiliary fields to calculate the finite and renormalized
power spectrum. Also, in Ref. \cite{c31}, by the Planck results (2013)
for scalar spectral index, we showed that the index of the Hankel
function $\mu$ lies in the range of $1.51\leq \mu \leq 1.53$, and
this important result stimulates us to move from the dS mode to
excited dS mode. Therefore in approximate de Sitter space-time,
according to (\ref{equ6}), the general solutions of the equations
of motion, including negative and positive frequency solutions, can
be given by \cite{c27}, \b \label{equ9}
u_{k}^{edS}\simeq\frac{A_{k}}{\sqrt{2k}}\Big(1-\frac{i}{k\eta}-\frac{1}{2}(\frac{i}{k\eta})^{2}\Big)e^{-ik\eta}\e
  $$+\frac{B_{k}}{\sqrt{2k}}\Big(1+\frac{i}{k\eta}-\frac{1}{2}(\frac{-i}{k\eta})^{2}\Big)e^{+ik\eta}.$$\\

It is necessary to emphasise that we consider the mode functions
(\ref{equ6}) for pure de Sitter space-time as the general exact
solution of (\ref{equ4}), while for approximate de Sitter
space-time, we consider excited-de Sitter mode functions
(\ref{equ9}) as the general approximate solutions of (\ref{equ4}).\\

\subsection{Power Spectrum with Excited-de Sitter Mode}
According to the usual definition used for matter density
perturbations, the power spectrum of the scalar field has
dimention $ k^{-3},$ so we will have \cite{c5}
\begin{equation} \label{equ10} p_{\varphi}(k)=\frac{|u_{k}(\eta)|^{2}}{a^{2}}. \end{equation}
By inserting mode function (8) in (10) and doing some
straightforward calculations, the power spectrum is obtained by
\cite{c27} \b
 p_{\varphi}(k)=(\frac{H^{2}}{2k^{3}})\Big(2+\frac{H^{2}}{4M_{*}^{2}}\Big),
\e where $ M_{*}=-Hk\eta_{*} $ and $ H\ll M_{*}<M_{P}$ is the
scale of new physics or the cutoff scale.

\section{Large-Scale Inhomogeneity and Anisotropy}
\subsection{Large-scale matter density perturbation}
So far, non-linear corrections due to applying an excited-de Sitter
mode have been reviewed. What we are interested in is
investigation of its effect on the large-angular CMB anisotropy or,
equivalently, the Sachs-Wolfe effect. After inflation, when density
perturbations re-enter the horizon, they do not grow appreciably
before matter-domination \cite{c1,c2,c3}. At this time pressure is
too large to allow increasing of density perturbations. In other
words, the density of radiation acts as a repulsive force and
suppresses the growth of inhomogeneities whose scales are smaller
than the horizon, while large-scale inhomogeneities remain unaffected
\cite{c1,c2,c3,c5}. So the horizon size at equality is an
important scale for investigating
structure formation.\\
After the equality epoch, in the matter domination era, density
perturbations are related to the gravitational potential via the
perturbed Poisson's equation in the short wave-length limit:
\cite{c2}
\begin{equation} \label{equ12} \Phi(k,a)=\frac{4\pi G \rho_{m}a^{2}}{k^{2}}\delta(k,a),\end{equation}
where $ 4\pi G \rho_{m}=(\frac{3\Omega_{m}}{2a^{2}})H_{0}^{2} $.
So the density perturbation is
\begin{equation} \label{equ13} \delta(k,a)=\frac{2k^{2}}{3\Omega_{m}H_{0}^{2}}\Phi(k,a).\end{equation}
The primordial potential formed during the inflationary epoch is
related to the gravitational potential after equality. On the other
hand, when the universe passes through equality, the potential on the
large scale drops by a factor of $ (9/10) $ \cite{c1, c2}. So in
this case, the relation between the two of them takes the
following form \cite{c2},
\begin{equation} \label{equ14} \Phi(k,a)=\frac{9}{10}\Phi_{p}(k)T(k)\frac{D_{1}(a)}{a}, \end{equation}
where $ \Phi_{p}(k)$ and $ T(k)$ are the primordial potential and
transfer function respectively. The transfer function determines
the modification of the fluctuation amplitudes (due to their evolution
through the horizon crossing) at different scales. $ D_{1}(a) $ is
called the growth function and describes the growth of the
perturbation amplitudes after the epoch of equality. Considering (13)
together with (14) yields
\begin{equation} \label{equ15} \delta(k,a)=\frac{3}{5}\frac{k^{2}}{\Omega_{m}H_{0}^{2}}\Phi_{p}(k)T(k)D_{1}(a). \end{equation}
So, the power spectrum of density perturbations after equality is
related to the power spectrum of the primordial potential by
\begin{equation} \label{equ16} P_{\delta}(k,a)=\frac{9}{25}\Big(\frac{k}{H_{0}}\Big)^{4}P_{\Phi_{p}}(k)T^{2}(k)\Big(\frac{D_{1}^{2}(a)}{\Omega_{m}^{2}}\Big)^{2}.\end{equation}
In the slow-roll approximation, by considering $
P_{\Phi_{p}}(k)=\frac{16\pi G}{9\epsilon}p_{\varphi}(k) $ and
(11), we have
\begin{equation} \label{equ17} P_{\Phi_{p}}=\frac{32\pi G}{9\epsilon}(\frac{H^{2}}{2k^{3}})(1+\frac{H^{2}}{8M_{*}^{2}}). \end{equation}
Now, this result can be used for finding the power spectrum of
matter density perturbations in the matter domination era as
\begin{equation} \label{equ18} P_{\delta}(k,a)=4\pi^{2} \delta_{H}^{2}T^{2}(k)\Big(\frac{D_{1}(a)}{D_{1}(a=1)}\Big)^{2}(\frac{k}{H_{0}^{4}})(1+\frac{H^{2}}{8M_{*}^{2}}), \end{equation}
where $ \delta_{H}^{2} $ is the Harrison-Zel'dovich-Peebles power
spectrum on the horizon scale. In the limit of very large scale,
the transfer function can be replaced by unity in (18), so the
matter perturbation power spectrum at present takes the following
form
\begin{equation} \label{equ19} P_{\delta}(k,a)=4\pi^{2} \delta_{H}^{2}(\frac{k}{H_{0}^{4}})(1+\frac{H^{2}}{8M_{*}^{2}}). \end{equation}\\
Also, in the regime of slow-roll inflation one can assume the Hubble
parameter is scale dependent as
\begin{equation} \label{equ20}\frac{H}{H_{*}}\sim (\frac{k}{k_{*}})^{-\epsilon},\end{equation}
where $H_{*}$ is the Hubble parameter evaluated when the
perturbation with scale $k_{*}$ leaves the horizon \cite{c34}.
Substituting (20) into relation (19) yields
\begin{equation} \label{equ21}P_{\delta}(k,a)=4\pi^{2} \delta_{H}^{2}(\frac{k}{H_{0}^{4}})[1+\frac{H_{*}^{2}}{8M_{*}^{2}}(\frac{k}{k_{*}})^{-\epsilon}]. \end{equation}
\subsection{Large Angular Scale CMB Anisotropy}
As mentioned previously, the angular scale which
corresponds to the horizon scale at recombination $ (\theta \sim
1^{o}) $ is an important scale as a dividing line between
large-scale and small-scale inhomogeneities. The temperature
fluctuations at large angular scales $ (\theta >> 1^{o}) $ are
induced by large-scale perturbations which are not affected by
photon pressure before equality. So observation of the temperature
fluctuations at these scales gives direct information about
the primordial power spectrum of density
perturbations.\\
In Section (III.A), we noticed that employment of the new
spectrum for calculating the matter density spectrum leads to the
scale dependent matter density power spectra. Now, we will use the
result obtained in (21) to calculate the power spectrum of temperature
fluctuations of CMB photons at large angular scale. If the CMB
anisotropy at this scale is denoted by $ {C_{\ell}}^{SW} $, it can
be read as \cite{c2} \b \label{equ22} {C_{\ell}}^{SW}\approx
\frac{\Omega_{m}^{2}H_{0}^{4}}{2\pi D_{1}^{2}(a=1)}
\int_{0}^{\infty}\frac{dk}{k}P_{\delta}(k)j_{\ell}^{2}(k(\eta_{0}-\eta_{rec})),
\e where $ \eta_{rec} $ is the conformal time at recombination. By
considering $ \eta_{0} >> \eta_{rec} $, one can neglect $
\eta_{rec} $ compared to $ \eta_{0} $ in (22). Moreover, in the
case of large-angular scale CMB anisotropy, the transfer function
can be set to unity. So, by substituting (21) in (22), $
{C_{\ell}}^{SW} $ takes the following form \b
\label{equ23}{C_{\ell}}^{SW}\approx \frac{2\pi
\delta_{H}^{2}\Omega_{m}^{2}}{D_{1}^{2}(a=1)}
\int_{0}^{\infty}\frac{dk}{k}j_{\ell}^{2}(k\eta_{0})[1+\frac{H_{*}^{2}}{8M_{*}^{2}}(\frac{k}{k_{*}})^{-\epsilon}].
\e\\
If we neglect the correction term, by considering $ X=k\eta_{0} $ and
with the help of the identity \cite{c35} \b
\label{equ24}\int_{0}^{\infty} dX X^{2}j_{\ell}^{2}(X)=2^{n-4}\pi
\frac{\Gamma(\ell+\frac{n}{2}-\frac{1}{2})\Gamma(3-n)}{\Gamma(\ell+\frac{5}{2}-\frac{n}{2})\Gamma^{2}(2-\frac{n}{2})},
\e the temperature fluctuation spectra for the CMB radiation at
large angular scales would be read as \b
\label{25}{C_{\ell}}^{SW}= \frac{\pi \Omega_{m}^{2}
}{2D_{1}^{2}(a=1)}\frac{\delta_{H}^{2}}{\ell(\ell+1)}. \e So the
quantity $ \ell(\ell+1){C_{\ell}}^{SW} $ is independent of $ \ell
$. Indeed, this is the reason that the CMBA power spectrum
is typically plotted in form of $ \ell(\ell+1){C_{\ell}}^{SW} $
versus $ \ell $. Now, we consider the excited dS mode instead of the pure
dS mode and after straightforward calculations, we obtain the
scale-dependent new result as \b \label{26}
{C_{1}}^{SW}\approx\frac{\pi\delta_{H}^{2}}{2}(\frac{\Omega_{m}}{D_{1}(a=1)})^{2}[\frac{1}{\ell(\ell+1)}\e
$$+\frac{\pi H_{*}^{2}}{2M_{*}^{2}}(\frac{\eta_{0}k_{*}}{2})^{\epsilon}(\frac{\Gamma(\ell-\frac{\epsilon}{2})\Gamma(2+\epsilon)}{\Gamma(\ell+2+\frac{\epsilon}{2})\Gamma^{2}(\frac{3}{2}-\frac{\epsilon}{2})})].$$\\
Since in most of the Sachs-Wolfe limit we have 
$\frac{\epsilon}{2}\ll \ell $, we can ignore $\epsilon$ in 
comparison to $\ell$ and consequently the final answer can be
written in the following form, \b \label{27}\ell(\ell+1)
{C_{\ell}}^{SW}\approx \frac{\pi \delta_{H}^{2}
}{2}(\frac{\Omega_{m}}{D_{1}(a=1)})^{2}[1+\e
$$+\frac{\pi H_{*}^{2}}{2M_{*}^{2}}(\frac{\eta_{0}k_{*}}{2})^{\epsilon}(\frac{\Gamma(2+\epsilon)}{\Gamma^{2}(\frac{3}{2}-\frac{\epsilon}{2})})].$$\\
The correction term on the right hand side of (27), which results
from the Sachs-Wolfe effect, shows that the quantity $ \ell(\ell+1)
{C_{\ell}}^{SW} $ depends on $ \epsilon $ and has a slight
deviation from the standard scale-invariant result(25).

\section{Conclusions}
In this paper, we have calculated the scale-dependency of large
angular scale CMB anisotropy resulting from an excited-dS mode as the
fundamental mode function during inflation. The result 
indicates that considering the new excited mode as an initial
quantum state for primeval fluctuations can affect the angular
scale invariance of CMB anisotropy spectra. In addition, the appearance of
a cutoff scale in the results shows that the effect of these modes should
be suppressed by some unknown energy scale, which must be higher
than the inflationary scale. This excited mode was prepared
essentially by expanding the Hankel function in the quantum
mode in de Sitter space-time to quadratic order of
$\frac{1}{k\eta}$ before quantization. This approach is similar to
performing the quantization at finite wavelength, rather than
fully in the ultraviolet (i.e. Bunch-Davies) limit. In fact,
taking into account the resent observational constraint together
with the result obtained in \cite{c31} motivates us to use quasi-de
Sitter curved space-time and excited mode functions. We have seen
that slight deviation of the Bunch-Davies mode leads to 
corrections in the primordial gravitational potential spectra and
in the large angular scales CMB anisotropy. The
result also shows that the final $\epsilon$-dependent correction term
is very tiny for low $\ell<200$ and close to the plateau result,
while for the limit $\ell\ll 200$, maybe due to dark energy
fluid evolution, the size of the correction term can be significant.
Also, from the Planck data, it is known that there might exist an
anomaly of the power asymmetry at $\ell < 50$, as analyzed in
\cite{c36, c37}. We plan to check the possible connection
between this anomaly and the present study. In our next study, we
plan to examine the accuracy of our approach observationally.

\noindent {\bf{Acknowledgements}}: This work has been supported by
the Islamic Azad University, Rasht Branch, Rasht, Iran.

\end{document}